\RequirePackage{ifpdf}
\newcommand{\commondocopts}{letterpaper,aps,prl,10pt,superscriptaddress,showpacs,floats,twocolumn}
\ifpdf
  \documentclass[\commondocopts,pdftex]{revtex4-1}
  \usepackage{cmap}
\else
  \documentclass[\commondocopts,ps2pdf]{revtex4-1}
\fi
\usepackage[latin1]{inputenc}
\usepackage[T1]{fontenc}

\newcommand{\titlestr}{Multi-electron transitions induced by neutron
  impact on helium}

\usepackage{placeins}
\usepackage{amsmath,amssymb}

\usepackage{graphicx,color}
\usepackage{hyperref}
\definecolor{rltred}{rgb}{0.75,0,0}
\definecolor{rltgreen}{rgb}{0,0.5,0}
\definecolor{rltblue}{rgb}{0,0,0.75}
\hypersetup{colorlinks,hypertexnames=true,pdfauthor={Matthias Liertzer, Johannes Feist, Stefan Nagele, Joachim Burgd\"orfer},pdftitle={\titlestr},pdfpagemode=UseNone,bookmarksopen=true,bookmarksnumbered=true,pdfhighlight=/I,linkcolor=rltred,citecolor=rltgreen,linktocpage=true}

\graphicspath{{plots/}}

\begin{document}
 
\newcommand{\Int}{\int\limits}
\newcommand{\IInt}{\iint\limits}
\newcommand{\IIInt}{\iiint\limits}
\newcommand{\IIIInt}{\iiiint\limits}
\newcommand{\ui}{\mathrm{i}}
\newcommand{\ue}{\mathrm{e}}
\newcommand{\Vol}{\operatorname{vol}}
\newcommand{\R}{\mathbb{R}}
\newcommand{\Z}{\mathbb{Z}}
\newcommand{\N}{\mathbb{N}}
\newcommand{\cA}{\mathcal{A}}
\newcommand{\E}{\mathcal{E}}
\newcommand{\setsep}{ \;\; | \;\;}

\newcommand{\COMMENT}[1]{\textcolor{rltred}{\textsc{\textbf{#1}}}}
\newcommand{\etal}{\emph{et~al.\@} }
\newcommand{\ie}{i.e., }
\newcommand{\Schro}{Schr\"o\-din\-ger }
\newcommand{\eg}{e.g.\@ }
\newcommand{\cf}{cf.\@ }

\newcommand{\expval}[1]{\langle#1\rangle}
\newcommand{\abs}[1]{\left|#1\right|}
\newcommand{\norm}[1]{\left\|#1\right\|}

\newcommand{\bra}[1]{\langle#1|}
\newcommand{\ket}[1]{|#1\rangle}
\newcommand{\braket}[2]{\langle#1|#2\rangle}

\newcommand{\doubD}{{\mathord{\buildrel{\lower3pt\hbox{$\scriptscriptstyle\leftrightarrow$}}\over {\bf D}}}}
\newcommand{\unitv}[1]{\mathbf{\hat{#1}}}

\renewcommand{\equationautorefname}{Eq.}
\renewcommand{\figureautorefname}{Fig.}

\newcommand{\refeq}[1]{\hyperref[#1]{\equationautorefname~(\ref*{#1})}}

\newcommand{\cvec}[1]{\mathbf{#1}}
\newcommand{\op}[1]{\mathrm{\hat{#1}}}
\newcommand{\vecop}[1]{\cvec{\hat{#1}}}
\newcommand{\eqcomma}{\,,\;\;}
\newcommand{\ed}{\,}

\newcommand{\dx}{\dd x}
\newcommand{\dt}{\dd t}
\newcommand{\dr}{\dd r}
\newcommand{\dw}{\dd\omega}
\newcommand{\dwb}{\dd\bar{\omega}}
\newcommand{\dW}{\dd\Omega}
\newcommand{\dE}{\dd E}
\newcommand{\dk}{\dd k}
\newcommand{\dd}{\mathrm{d}}
\newcommand{\Dt}{\Delta t}
\newcommand{\hw}{\hbar\omega}

\newcommand{\sube}{{\mathrm{e}}}
\newcommand{\subn}{{\mathrm{n}}}

\newcommand{\csph}{{\mathcal{Y}}}
\newcommand{\sph}[2]{{\mathrm{Y}_{#2}^{(#1)}}}
\newcommand{\sphcmplx}[2]{{\mathrm{Y}_{#2}^{*(#1)}}}
\newcommand{\rensph}[2]{{\mathrm{C}_{#2}^{(#1)}}}
\newcommand{\renredsph}[1]{{\mathrm{C}^{(#1)}}}
\newcommand{\submax}{\mathrm{max}}
\newcommand{\Lmax}{L_\submax}
\newcommand{\lmax}{l_\submax}
\newcommand{\lonemax}{l_{1,\submax}}
\newcommand{\ltwomax}{l_{2,\submax}}

\newcommand{\cmfs}{\,\text{cm}^4\text{s}}
\newcommand{\ev}{\,\text{eV}}
\newcommand{\eV}{\ev}
\newcommand{\au}{\,\text{a.u.}}
\newcommand{\nm}{\,\text{nm}}
\newcommand{\He}{\text{He}}
\newcommand{\Hep}{\text{He}^{+}}
\newcommand{\Wcm}{\,\text{W}/\text{cm}^2}
\newcommand{\as}{\,\text{as}}
\newcommand{\fs}{\,\text{fs}}

\newcommand{\ti}{{t_\text{(i)}}}
\newcommand{\tii}{{t_\text{(ii)}}}
\newcommand{\tiii}{{t_\text{(iii)}}}
\newcommand{\tcor}{{t_\text{cor}}}
\newcommand{\Tp}{{T_{\text{p}}}}
\newcommand{\Teff}{{T_{\text{eff}}}}
\newcommand{\Tramp}{{T_{\text{ramp}}}}
\newcommand{\Tfull}{{T_{\text{full}}}}
\newcommand{\Ene}{{E_\text{ne}}}
\newcommand{\Eexc}{{E_\text{exc}}}
\newcommand{\Etot}{{E_\text{tot}}}

\newcommand{\level}[3]{{}^{#1}{#2}^{\textrm{#3}}}

\newcommand{\redmate}[3]{\left<#1\left|\left|#2\right|\right|#3\right>}

\newcommand{\CGC}[6]{\left<\left. #1 #2 #3 #4\vphantom{#5 #6} \right| \vphantom{#1 #2 #3 #4} #5 #6 \right>}

\newcommand{\CG}[6]{\left[ \! \begin{array}{ccc}
#1 & #2 & #3 \\
#4 & #5 & #6
\end{array} \! \right]}

\newcommand{\Threej}[6]{\left( \!\! \begin{array}{ccc}
#1 & #2 & #3 \\
#4 & #5 & #6
\end{array} \!\! \right)}

\newcommand{\Sixj}[6]{\left\{ \!\! \begin{array}{ccc}
#1 & #2 & #3 \\
#4 & #5 & #6
\end{array} \!\! \right\}}

\title{\titlestr}

\author{M.~Liertzer}
\email{matthias.liertzer@tuwien.ac.at}
\affiliation{Institute for Theoretical Physics, 
             Vienna University of Technology, 1040 Vienna, Austria, EU}

\author{J.~Feist}
\affiliation{ITAMP, 
            Harvard-Smithsonian Center for Astrophysics, Cambridge, Massachusetts 02138, USA}
\affiliation{Institute for Theoretical Physics, 
             Vienna University of Technology, 1040 Vienna, Austria, EU}

\author{S.~Nagele}
\affiliation{Institute for Theoretical Physics, 
             Vienna University of Technology, 1040 Vienna, Austria,
             EU}

\author{J.~Burgd\"orfer}
\affiliation{Institute for Theoretical Physics, 
             Vienna University of Technology, 1040 Vienna, Austria, EU}

\date{\today}

\begin{abstract}
  We explore excitation and ionization by neutron impact as a novel
  tool for the investigation of electron-electron correlations in
  helium. We present single and double ionization spectra calculated
  in accurate numerical ab-initio simulations for incoming neutrons
  with kinetic energies of up to 150\,keV. The resulting electron
  spectra are found to be fundamentally different from photoionization
  or charged particle impact due to the intrinsic many-body character
  of the interaction. In particular, doubly excited resonances that
  are strongly suppressed in electron or photon impact become
  prominent. The ratio of double to single ionization is found to
  differ significantly from those of photon and charged particle
  impact.
\end{abstract}
\maketitle

Spectroscopic studies of atoms, molecules, and solids rely on the well
established excitation processes such as photoabsorption and
charged-particle impact. The underlying dynamical processes are theoretically well-understood
within the framework of linear response of the system to the external
probe. The observables accessible by these probes are, however,
limited by either exact selection rules or approximate ``propensity''
rules. For example, photoabsorption spectroscopy is strongly dominated
by dipole-allowed transitions. In charged-particle impact, higher
multipole transitions are allowed but are typically suppressed in
``soft'' collisions with small momentum transfers. Moreover, the
long-range Coulomb interactions between the probing particle and the
excited system may distort the excitation and ionization to be
extracted by ``post-collision'' interactions which are typically
beyond lowest-order perturbation (LOP) theory underlying linear
response.

Photon and charged particle interactions have in common that the LOP
interaction is strictly a one-body operator. The point of departure of
our present study is the observation that neutron impact gives rise to
intrinsic many-body interactions in the electronic system
\cite{Ber2001}. The underlying idea is that neutron scattering
at the atomic nucleus gives rise to a sudden ``kick''. In the frame of
the atom, this results in a simultaneous momentum boost for \emph{all}
electrons, effectively causing a true \emph{many-body} transition
which can efficiently lead to multiple excitation and ionization of the
atom. In this Letter we theoretically investigate the neutron-impact
ionization of helium atoms. Helium is the prototypical case of a
strongly correlated system \cite{TanRicRos2000} in both the ground state and doubly excited
resonances which can be treated exactly by numerical ab initio
calculations (\cf \eg \cite{KheBra1996,ParSmyTay1998,MccBaeRes2004,FouAntPir2008}). It thus serves as testing ground for the study of
electron correlation and multi-electron effects. We show that neutron
impact leads to a very broad energy distribution in the final states
including double ionization and, furthermore, that it can efficiently
produce doubly excited states that are disfavored by other probing
agents.

\begin{figure}[bp]
  \centering
  \includegraphics{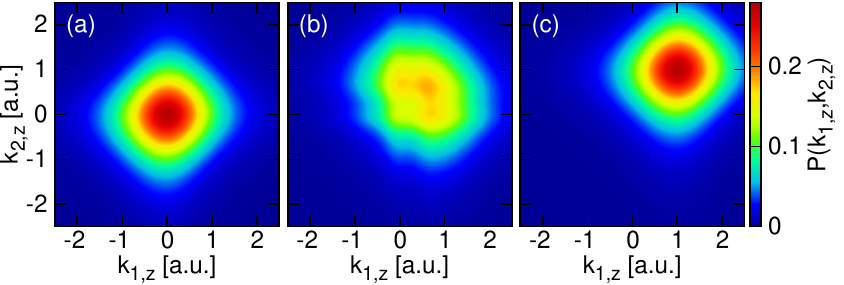}
  \caption{Projected two-electron momentum distribution $P(k_{1,z},k_{2,z})$ for (a) the
    ground state of helium, (b) the ground state wave function boosted by the one-body boost
    operator $B_\mathrm{1B}$ with a momentum transfer $\Delta p_e = 1.0\au$, (c) boosted
    by the collective boost operator $B_\mathrm{c}$ (\autoref{eq:Bc}) with identical $\Delta p_e$ (see text).}
  \label{fig:boostwf}
\end{figure}

We assume that the only interaction in the neutron-helium collision is
quasi-elastic scattering between the neutron and the nucleus, mediated
by the strong nuclear force.  The contributions of magnetic
interactions of the neutron with the electronic and nuclear magnetic
moment are small enough to be safely neglected \cite{Sea1986}. Neutron
energies are kept sufficiently low in order to exclude any inelastic
nuclear processes. The duration of the neutron-nucleus scattering event
is much shorter than the typical time scale of the dynamics of
electrons bound to the nucleus ($\sim$ attoseconds) currently probed using ultrashort light pulses \cite{HenKieSpi2001,DreHenKie2001,KraIva2009}. 
Electronic transitions can therefore be described by an impulse or ``sudden''
approximation. Accordingly, the transition amplitude for quasi-elastic
scattering of the neutron accompanied by an electronic transition $i
\to f$ is given by
\begin{equation}
  t_{if}(\Delta \vec{p}_{nuc}) \approx t^{el}_{nuc}(\Delta \vec{p}_{nuc})
  \cdot t_{i,f}^e(\Delta\vec{p}_e), \label{eq:transition}
\end{equation}
where $t^{el}_{nuc}$ is the transition amplitude for elastic nuclear
scattering with momentum transfer $\Delta \vec{p}_{nuc} = \vec{k}_f -
\vec{k}_i$ and $t_{i,f}^e$ is the matrix element of the collective
boost operator
\begin{equation}
  t^e_{i,f}(\Delta\vec{p}_e) = \bra{\Psi_f} \exp[i \Delta \vec{p}_e\cdot
  (\vec{r}_1 + \vec{r}_2)] \ket{\Psi_i} \label{eq:transition_electronic}
\end{equation}
with
\begin{equation}
  \Delta\vec{p}_e = - \frac{\Delta\vec{p}_{nuc}}{M_\alpha + 2} \label{eq:momtransfer}
\end{equation}
and $M_\alpha$ the mass of the $\alpha$ particle in atomic
units. Taylor expansion of the collective boost operator
\begin{align}
  \label{eq:Bc}
  B_\mathrm{c}(\Delta\vec{p}_e) &= \exp[i \Delta \vec{p}_e \cdot(\vec{r}_1 +
  \vec{r}_2)]\\ &\approx 1 + i\Delta\vec{p}_e\cdot(\vec{r_1}+\vec{r_2}) -
  \frac{1}{2}[\Delta\vec{p}_e\cdot(\vec{r}_1+\vec{r}_2)]^2 \label{eq:boost_taylor}
\end{align}
shows that while, to first order in $\Delta\Vec{p}_e$, the electronic
transition matrix element is equivalent to that of the one-body
operator from photoabsorption or the Bethe-Born limit of soft
charged-particle collisions, all higher-order terms represent a true
many-body transition structurally different from photon or
charged-particle interactions. Application of the collective boost to
the exact helium ground state (\autoref{fig:boostwf}) leads to a
correlated displacement of the projected two-electron momentum
distribution unlike the one-body boost operator,
$B_\mathrm{1B}(\Delta\vec{p}_e) = \sum_{i=1}^{N}
\exp(i\Delta\vec{p}_e\cdot\vec{r}_i)$ governing, for example, Compton
scattering or charged particle impact on an $N$-electron atom.  This
property plays a key role in accessing states blocked by parity or
propensity rules.

Differential cross sections for electronic
inelastic processes accompanied by quasi-elastic neutron-alpha
particle scattering are given by
\begin{equation}  
  \frac{\dd\sigma_{i{\to}f}}{\dd\Omega}(\Delta\vec{p}_{nuc}) = \frac{k_f}{k_i}
  \frac{\dd\sigma_{el}}{\dd\Omega}(\Delta\vec{p}_{nuc}) |t_{i{\to}f}^e(\Delta p_e)|^2
  \label{eq:diffcross}
\end{equation}
with $k_f = \sqrt{k_i^2 - 2\mu Q_I}$, $Q_I=E_f^e-E_i^e$ the internal excitation
energy, and $\mu$ the reduced mass of the n-He system.

For the nuclear elastic scattering cross section
$\frac{\dd\sigma_{el}}{\dd\Omega}$ we use the tabulated data from
\footnote{The angular dependent differential cross section has been
  obtained from the sigma database at the
  \href{http://www.nndc.bnl.gov/}{National Nuclear Database Center
    (NNDC)}.}.  For the electronic degrees of freedom in helium we
perform full ab-initio calculations by solving the six-dimensional
time-independent \Schro equation (five-dimensional after exploiting
cylindrical symmetry) including all interparticle interactions. In our
computational approach we employ a close-coupling scheme, in which the
angular variables are expanded in coupled spherical harmonics (with
total angular momentum up to $L_\submax=7$, and individual electron
angular momenta up to $l_\submax = 9$).  For the discretization of the
radial components we use a finite element discrete variable
representation (FEDVR) \cite{FeiNagPaz2008,FeiNagPaz2009}.  The
momentum boost operator Eq.~\eqref{eq:boost_taylor} is implemented
using a short iterative Lanczos algorithm (SIL)
\cite{ParkLight86}. For the extraction of transition amplitudes, the
direct projection onto final states would be most desirable but
unfeasible as exact three-body Coulomb continuum states are not
known. We therefore make use of an alternative approach
\cite{PalMccRes2007,MccBaeRes2004}, 
in which the Fourier transform of
the boosted wave packet is effectively calculated by solving
the inhomogeneous linear system
\begin{equation}
  (E-H)\ket{\Psi_\textrm{sc}(E)} = B_\mathrm{c}(\Delta\vec{p}_e)\ket{\Psi_i},
\end{equation}
where $\Psi_\textrm{sc}(E)$ is the scattered wave function in the
(time-independent) energy domain.  Outgoing boundary conditions are
enforced by an exterior complex scaling (ECS) transformation for each
of the radial coordinates.  For the calculations presented in this
Letter we chose an exterior scaling radius of $120\au$ and an overall
box size of up to $180\au$\,. The ejected single and double ionization
amplitudes can then be extracted from the scattering amplitude by
means of a surface integral within the non-scaled part of the grid
\cite{PalMccRes2007}.

\begin{figure}[tb]
  \centering
  \includegraphics{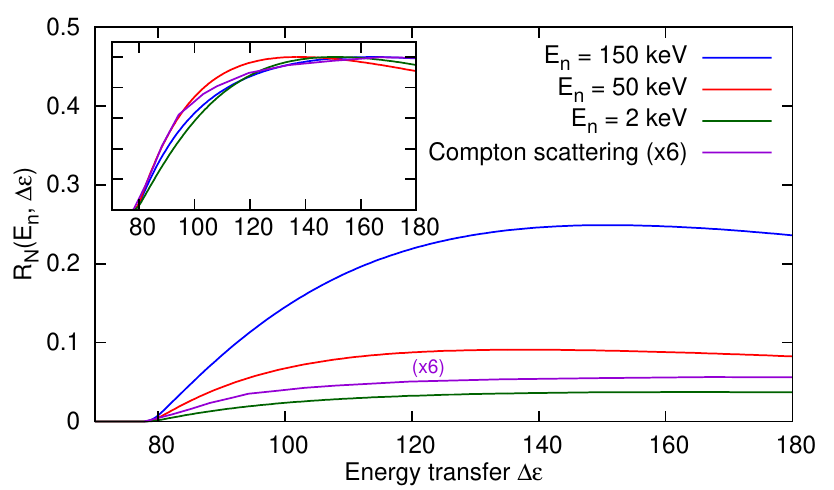}
  \caption{(Color online) Ratio of double to single ionization for
    neutron-impact ionization $R_\mathrm{N}(\Delta \varepsilon)$ as a
    function of energy transfer $\Delta \varepsilon$ compared to the
    corresponding ratio $R_\mathrm{C}$ for Compton scattering
    \cite{BurQiuWan1997} multiplied by a factor $6$ for visibility. 
    Inset: The same ratios normalized to their respective maxima.}
 \label{fig:totprobs}
\end{figure}
\begin{figure}
  \centering
  \includegraphics{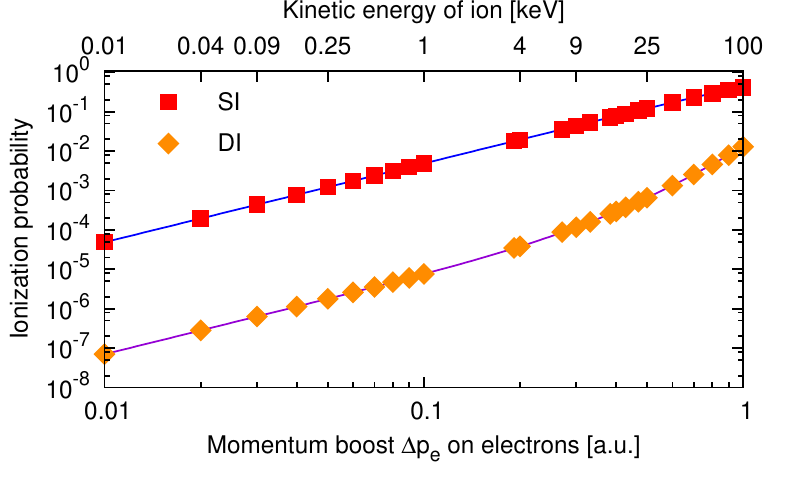}
  \caption{(Color online) Single (SI) and double ionization (DI)
    probabilities of helium as a function of the momentum boost $\Delta p_e$ for
    the electrons. The corresponding recoil energy of the kicked
    helium nucleus is given on the upper abscissa.}
  \label{fig:ionprob_pe}
\end{figure}

The most frequently studied quantity in double ionization of helium, a
paradigm for studying the role of electron correlation, is the ratio of double to
single ionization $R$. This ratio has been probed for both charged
particle impact and photon impact over a wide range of energies, both
experimentally and theoretically \cite{*[{For a review, see }] [{}] McGBerBar1995}. 
For photon impact, photoabsorption as well as Compton scattering have been studied
\cite{ByrJoa1967,Abe1970,DalSad1992,AndBur1993,AndBur1994,SurLogPra1994,BurQiuWan1997}. Compton 
scattering involving a neutral projectile and the one-body boost
operator as transition operator is expected to bear closest
resemblance to the present case of neutrons. Significant differences
are, however, expected, as for neutrons the collective boost rather
than the one-body boost controls the transition and, moreover,
different regions in the energy transfer ($\Delta \varepsilon)$ -
momentum transfer ($\Delta p_e$) plane are sampled. The ratio for
neutrons, $R_\mathrm{N}(\Delta \varepsilon$), differential in energy
transfer $\Delta \varepsilon$ to the electronic system, qualitatively
resembles the calculated $R_\mathrm{C}(\Delta \varepsilon)$ for
Compton scattering near threshold (inset \autoref{fig:totprobs}). Its
absolute magnitude is, however, strongly enhanced by factors up to 25
depending on the kinetic energy of the incident neutron
(\autoref{fig:totprobs}).
\begin{figure}
  \centering
  \includegraphics{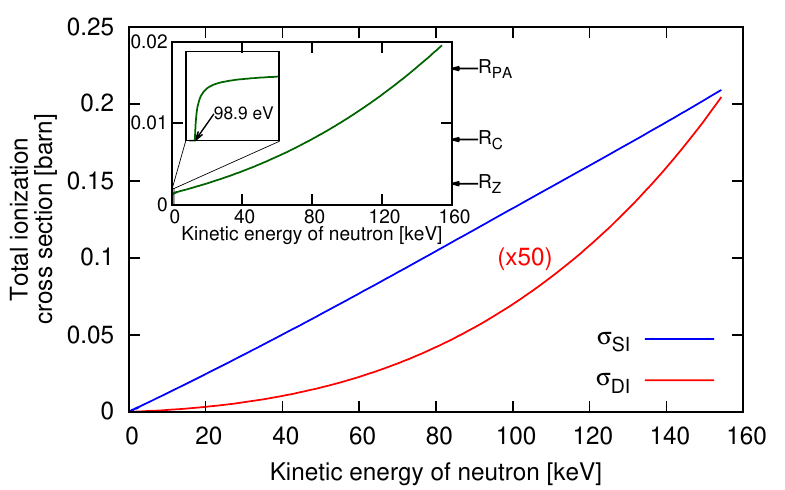}

  \caption{(Color online) Absolute integrated double
    ($\sigma_{\mathrm{DI}}$) and single ionization
    ($\sigma_{\mathrm{SI}}$) cross section by neutron impact as a
    function of the neutron kinetic energy. Inset: Energy dependence
    of $R_\mathrm{N} = \sigma_{\mathrm{DI}} / \sigma_{\mathrm{SI}}$
    with magnification of threshold region. The nonrelativistic
    high-energy limits for photoabsorption $R_\mathrm{PA}$, Compton
    scattering $R_\mathrm{C}$, and charged particle impact
    $R_\mathrm{C}$ are shown for comparison.}
  \label{fig:totscatt}
\end{figure}
The most dramatic difference (\autoref{fig:ionprob_pe}) occurs for large momentum transfers due
to the non-linear dependence of the boost operator on $\Delta p_e$. 
In the limit $\Delta p_e \to \infty$, or
more precisely when the momentum transfer is large compared to the
width of the momentum distribution of the initial state, $\Delta p_e
\gg \langle p_e^2\rangle^{\frac{1}{2}}$, the ratio diverges, as the
strongly displaced momentum distribution (\autoref{fig:boostwf}) will effectively
cease to overlap with bound states and double ionization
dominates. 

Most easily accessible in experimental investigations is the ratio
$R_\mathrm{N}$ of total double to single ionization cross section
(Inset \autoref{fig:totscatt}) resulting from integration of
Eq.~\eqref{eq:diffcross} over all accessible final states in the
$\Delta \varepsilon$-$\Delta p_e$ plane as a function of the kinetic
energy of the incident neutron. With increasing neutron energy the
ratio $R_N$ increases polynomially $(\propto a_1 E_\mathrm{N} + a_2
E_\mathrm{N}^2 + \cdots)$ with the neutron energy and eventually
surpasses the well known (non-relativistic) high-energy limits for
photoabsorption (1.66\%)
\cite{ByrJoa1967,Abe1970,DalSad1992,AndBur1993}, Compton scattering
(0.8\%) \cite{AndBur1994,SurLogPra1994,BurQiuWan1997} and charged-particle impact (0.26\%)
\cite{AndHveKnu1987,McGBerBar1995}. The reason is that the He nucleus suddenly
``disappears'' from the electronic charge cloud resulting in a high
probability for double ionization.

\begin{figure}
  \centering
  \includegraphics[width=\linewidth]{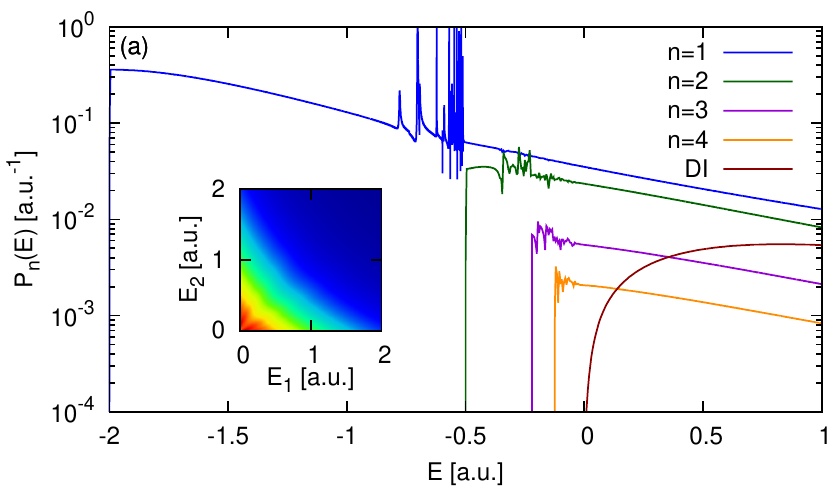}
  \includegraphics[width=\linewidth]{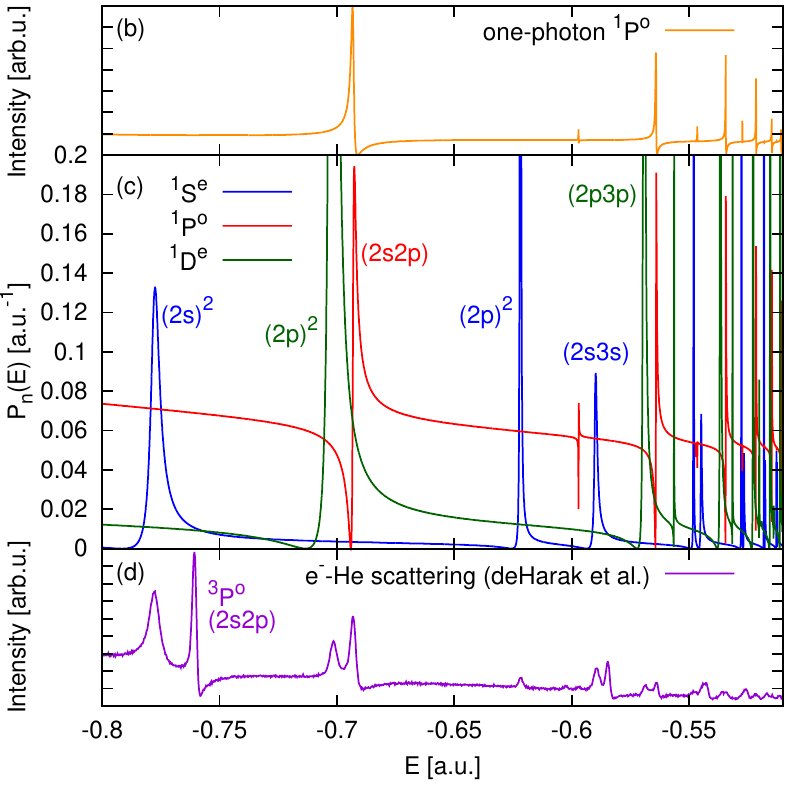}
  \caption{(Color online) (a) Electron spectrum as a function of final
    energy for single and double ionization for a kick strength of
    $\Delta p_e=1\au$. The energy of the ejected electron in the case
    of SI is given by $E+\frac{2}{n^2}$ with the remaining ionized
    helium being excited to state $n$. In the case of double
    ionization the probability for ejecting electrons with the sum of
    the individual electron energies equivalent to $E$ is plotted. The
    full two-electron energy distribution is plotted in the inset of
    (a).  A close-up of Fano resonances in the $n=1$ channel for
    different final symmetries is shown in (c). The first few
    resonances are labeled by approximate independent-particle
    configurations.  For comparison, the one-photon spectrum (b,
    calculated) and an ejected electron spectrum (d) from e${}^-$-He
    scattering experiments performed by deHarak \etal \cite{*[{The
        spectrum, originally published in }] [{, was taken at an angle
        of $120^\circ$ with respect to a 75 eV incident electron
        beam.}] HarChiMar2006} are shown. \label{fig:energy_spectra}}
\end{figure}

A more sensitive probe of the momentum shift of the two-electron
momentum distribution by the collective boost is the energy spectrum of
the ejected electrons in single and double ionization
(\autoref{fig:energy_spectra}). The impulsive momentum transfer leads to
a broad-band excitation (the upper cut-off due to the finite nuclear
collision time $\sim 1 / t_{coll}$ lies well beyond the spectral
range shown in \autoref{fig:energy_spectra}) resulting in a large
number of doubly excited resonances embedded in the single ionization
continuum.
A zoom into the electron energy spectrum just below the $n=2$
threshold [\autoref{fig:energy_spectra}(c)]
shows the multitude of Beutler-Fano resonances of different
symmetries. Note that doubly excited states are not properly
identified by the usual independent-particle labels, but require
collective quantum numbers (\cf\cite{TanRicRos2000} and references
therein).  However, for brevity, we use the traditional but imprecise
labels $(nl\,n'l')$ to describe the first few doubly excited states.
The background from direct single ionization into the continuum is
only strong in the channel with $\level1Po$ final symmetry, while it
is suppressed in the other channels. This is a clear signature of the
different dominant terms in the transition operator for different
symmetries: in $\level1Po$, the first-order (one-body) part dominates,
which couples efficiently to the single continuum, but only weakly to
doubly excited states. In $\level1Se$ and $\level1De$, the dominant
part of the boost operator is the second-order two-body term. The
latter couples the initial ground state more efficiently to the
quasi-bound doubly excited states than to the single ionization
continuum. This is best seen in the $(2p)^2$ (both in the ${}^1S^e$ as
well as in the ${}^1D^e$ channel) and $(2p3p)$ doubly excited states
which feature the largest cross section. By contrast, these
transitions are strongly forbidden in photoabsorption driven by the
dipole operator (first term in \autoref{eq:boost_taylor}). Exciting
those resonances by photons would require a two-photon absorption
process triggered by an intense beam with well-tuned frequencies,
in reach with free-electron lasers \cite{MosJiaFou2007,SorWelBob2007}.
Even within the dipole-allowed
$\level1Po$ spectrum neutron-impact ionization leads to a marked
modification of the Beutler-Fano resonance profiles
\cite{Beutler35,Fano35,*Fano61} compared to photoabsorption
[\autoref{fig:energy_spectra}(b),(c)]. The latter is a signature of
interference between the first and third-order terms in
\autoref{eq:boost_taylor}.

It is instructive to compare the neutron-impact induced spectrum with
the corresponding spectrum for electron impact
[\autoref{fig:energy_spectra}(d), taken from
\cite{HarChiMar2006}]. While for charged-particle collisions higher
multipole transitions become allowed, the propensity for excitation of
resonances of different symmetry are markedly different.  Considering,
for example, the first two doubly excited resonances in the ${}^1S^e$
channel,
the $(2p)^2$ state is much stronger excited for neutron impact
ionization than for electron scattering.  This is in contrast to the
$(2s)^2$ doubly excited state, which is present in both excitation
processes. This difference can be explained by specific electron
correlation effects present in these doubly excited states. It has
been shown \cite{SinHer1973} that a major difference between the two
states lies in the expectation value of the angle $\theta_{12}$
between the two electrons. For the $(2s)^2$ state the electrons are
more likely situated opposite to each other whereas in the $(2p)^2$
case they have a tendency to be located on the same side of the
nucleus. For the quasi-instantaneous neutron kick it is suggestive
that both electrons will be pushed to the same side of the nucleus and
will thus have significant overlap with this class of resonances. This
behavior is less likely for excitation by an incoming electron which
interacts with the bound electrons via the long-ranged Coulomb force
and gives rise to transition matrix elements containing the one-body
boost operator. The strong excitation can thus be directly attributed
to the effective many-body nature of the neutron kick. In contrast to
neutron impact, the collision with an incoming electron can also
access ${}^3P^o$ states due to spin exchange processes, which can be
seen in \autoref{fig:energy_spectra}(d) for the ${}^3P^o (2s2p)$
state.

Analogous processes induced by neutron impact are of interest
also in larger systems, \eg in molecules, where they result in the
excitation of auto-detaching states and opening of dissociative
channels.  In solids, they represent the key processes underlying
electronically induced radiation damage triggered by energetic
neutrons subsequent to knocking nuclei from their lattice positions.

In conclusion, we have shown that neutron-impact ionization could
serve as a novel tool to probe correlated electronic dynamics in
many-body electron systems, specifically in helium. Key is the true
many-body nature of the correlated boost operator which allows
transitions that are either strictly forbidden or strongly suppressed
in either photoabsorption or charged-particle excitation. Doubly
excited resonances become prominent that are otherwise only barely
visible. The ratio of double to single ionization by neutrons,
$R_\mathrm{N}$, is another benchmark for the underlying differences of
the ionization process. The predicted ratios significantly differ from
those for photoabsorption, Compton scattering, and charged-particle
collisions. With the availability of high-intensity neutron sources,
the observation of these processes under well-characterized
single-collision conditions may come into reach.

\begin{acknowledgments}
  We thank B.\ deHarak for providing us with the data of the
  e${}^-$-He scattering measurements. S.N.\ and J.B.\ acknowledge
  support by the FWF-Austria, SFB-041 VICOM and P23359-N16. J.F.\
  acknowledges support from the NSF through a grant to ITAMP. M.L.\
  acknowledges funding by the Vienna Science and Technology Fund
  (WWTF) through project MA09-030.  The computational results have
  been achieved using the Vienna Scientific Cluster and NSF
  TeraGrid/XSEDE resources provided by NICS and TACC under grant
  TG-PHY090031.
\end{acknowledgments}

\FloatBarrier


\begin{thebibliography}{31}%
\makeatletter
\providecommand \@ifxundefined [1]{%
 \@ifx{#1\undefined}
}%
\providecommand \@ifnum [1]{%
 \ifnum #1\expandafter \@firstoftwo
 \else \expandafter \@secondoftwo
 \fi
}%
\providecommand \@ifx [1]{%
 \ifx #1\expandafter \@firstoftwo
 \else \expandafter \@secondoftwo
 \fi
}%
\providecommand \natexlab [1]{#1}%
\providecommand \enquote  [1]{``#1''}%
\providecommand \bibnamefont  [1]{#1}%
\providecommand \bibfnamefont [1]{#1}%
\providecommand \citenamefont [1]{#1}%
\providecommand \href@noop [0]{\@secondoftwo}%
\providecommand \href [0]{\begingroup \@sanitize@url \@href}%
\providecommand \@href[1]{\@@startlink{#1}\@@href}%
\providecommand \@@href[1]{\endgroup#1\@@endlink}%
\providecommand \@sanitize@url [0]{\catcode `\\12\catcode `\$12\catcode
  `\&12\catcode `\#12\catcode `\^12\catcode `\_12\catcode `\%12\relax}%
\providecommand \@@startlink[1]{}%
\providecommand \@@endlink[0]{}%
\providecommand \url  [0]{\begingroup\@sanitize@url \@url }%
\providecommand \@url [1]{\endgroup\@href {#1}{\urlprefix }}%
\providecommand \urlprefix  [0]{URL }%
\providecommand \Eprint [0]{\href }%
\providecommand \doibase [0]{http://dx.doi.org/}%
\providecommand \selectlanguage [0]{\@gobble}%
\providecommand \bibinfo  [0]{\@secondoftwo}%
\providecommand \bibfield  [0]{\@secondoftwo}%
\providecommand \translation [1]{[#1]}%
\providecommand \BibitemOpen [0]{}%
\providecommand \bibitemStop [0]{}%
\providecommand \bibitemNoStop [0]{.\EOS\space}%
\providecommand \EOS [0]{\spacefactor3000\relax}%
\providecommand \BibitemShut  [1]{\csname bibitem#1\endcsname}%
\let\auto@bib@innerbib\@empty
%</preamble>
\bibitem [{\citenamefont {Berakdar}(2002)}]{Ber2001}%
  \BibitemOpen
  \bibfield  {author} {\bibinfo {author} {\bibfnamefont {J.}~\bibnamefont
  {Berakdar}},\ }\href {\doibase 10.1088/0953-4075/35/1/105} {\bibfield
  {journal} {\bibinfo  {journal} {J. Phys. B}\ }\textbf {\bibinfo {volume}
  {35}},\ \bibinfo {pages} {L31} (\bibinfo {year} {2002})}\BibitemShut
  {NoStop}%
\bibitem [{\citenamefont {Tanner}\ \emph {et~al.}(2000)\citenamefont {Tanner},
  \citenamefont {Richter},\ and\ \citenamefont {Rost}}]{TanRicRos2000}%
  \BibitemOpen
  \bibfield  {author} {\bibinfo {author} {\bibfnamefont {G.}~\bibnamefont
  {Tanner}}, \bibinfo {author} {\bibfnamefont {K.}~\bibnamefont {Richter}}, \
  and\ \bibinfo {author} {\bibfnamefont {J.~M.}\ \bibnamefont {Rost}},\ }\href
  {\doibase 10.1103/RevModPhys.72.497} {\bibfield  {journal} {\bibinfo
  {journal} {Rev. Mod. Phys}\ }\textbf {\bibinfo {volume} {72}},\ \bibinfo
  {pages} {497} (\bibinfo {year} {2000})}\BibitemShut {NoStop}%
\bibitem [{\citenamefont {Kheifets}\ and\ \citenamefont
  {Bray}(1996)}]{KheBra1996}%
  \BibitemOpen
  \bibfield  {author} {\bibinfo {author} {\bibfnamefont {A.~S.}\ \bibnamefont
  {Kheifets}}\ and\ \bibinfo {author} {\bibfnamefont {I.}~\bibnamefont
  {Bray}},\ }\href {\doibase 10.1103/PhysRevA.54.R995} {\bibfield  {journal}
  {\bibinfo  {journal} {Phys. Rev. A}\ }\textbf {\bibinfo {volume} {54}},\
  \bibinfo {pages} {R995} (\bibinfo {year} {1996})}\BibitemShut {NoStop}%
\bibitem [{\citenamefont {Parker}\ \emph {et~al.}(1998)\citenamefont {Parker},
  \citenamefont {Smyth},\ and\ \citenamefont {Taylor}}]{ParSmyTay1998}%
  \BibitemOpen
  \bibfield  {author} {\bibinfo {author} {\bibfnamefont {J.~S.}\ \bibnamefont
  {Parker}}, \bibinfo {author} {\bibfnamefont {E.~S.}\ \bibnamefont {Smyth}}, \
  and\ \bibinfo {author} {\bibfnamefont {K.~T.}\ \bibnamefont {Taylor}},\
  }\href {\doibase 10.1088/0953-4075/31/14/001} {\bibfield  {journal} {\bibinfo
   {journal} {J. Phys. B}\ }\textbf {\bibinfo {volume} {31}},\ \bibinfo {pages}
  {L571} (\bibinfo {year} {1998})}\BibitemShut {NoStop}%
\bibitem [{\citenamefont {McCurdy}\ \emph {et~al.}(2004)\citenamefont
  {McCurdy}, \citenamefont {Baertschy},\ and\ \citenamefont
  {Rescigno}}]{MccBaeRes2004}%
  \BibitemOpen
  \bibfield  {author} {\bibinfo {author} {\bibfnamefont {C.~W.}\ \bibnamefont
  {McCurdy}}, \bibinfo {author} {\bibfnamefont {M.}~\bibnamefont {Baertschy}},
  \ and\ \bibinfo {author} {\bibfnamefont {T.~N.}\ \bibnamefont {Rescigno}},\
  }\href {\doibase 10.1088/0953-4075/37/17/R01} {\bibfield  {journal} {\bibinfo
   {journal} {J. Phys. B}\ }\textbf {\bibinfo {volume} {37}},\ \bibinfo {pages}
  {R137} (\bibinfo {year} {2004})}\BibitemShut {NoStop}%
\bibitem [{\citenamefont {Foumouo}\ \emph {et~al.}(2008)\citenamefont
  {Foumouo}, \citenamefont {Antoine}, \citenamefont {Piraux}, \citenamefont
  {Malegat}, \citenamefont {Bachau},\ and\ \citenamefont
  {Shakeshaft}}]{FouAntPir2008}%
  \BibitemOpen
  \bibfield  {author} {\bibinfo {author} {\bibfnamefont {E.}~\bibnamefont
  {Foumouo}}, \bibinfo {author} {\bibfnamefont {P.}~\bibnamefont {Antoine}},
  \bibinfo {author} {\bibfnamefont {B.}~\bibnamefont {Piraux}}, \bibinfo
  {author} {\bibfnamefont {L.}~\bibnamefont {Malegat}}, \bibinfo {author}
  {\bibfnamefont {H.}~\bibnamefont {Bachau}}, \ and\ \bibinfo {author}
  {\bibfnamefont {R.}~\bibnamefont {Shakeshaft}},\ }\href {\doibase
  10.1088/0953-4075/41/5/051001} {\bibfield  {journal} {\bibinfo  {journal} {J.
  Phys. B}\ }\textbf {\bibinfo {volume} {41}},\ \bibinfo {pages} {051001}
  (\bibinfo {year} {2008})}\BibitemShut {NoStop}%
\bibitem [{\citenamefont {Sears}(1986)}]{Sea1986}%
  \BibitemOpen
  \bibfield  {author} {\bibinfo {author} {\bibfnamefont {V.}~\bibnamefont
  {Sears}},\ }\href {\doibase 10.1016/0370-1573(86)90129-8} {\bibfield
  {journal} {\bibinfo  {journal} {Physics Reports}\ }\textbf {\bibinfo {volume}
  {141}},\ \bibinfo {pages} {281} (\bibinfo {year} {1986})}\BibitemShut
  {NoStop}%
\bibitem [{\citenamefont {Hentschel}\ \emph {et~al.}(2001)\citenamefont
  {Hentschel}, \citenamefont {Kienberger}, \citenamefont {Spielmann},
  \citenamefont {Reider}, \citenamefont {Milosevic}, \citenamefont {Brabec},
  \citenamefont {Corkum}, \citenamefont {Heinzmann}, \citenamefont {Drescher},\
  and\ \citenamefont {Krausz}}]{HenKieSpi2001}%
  \BibitemOpen
  \bibfield  {author} {\bibinfo {author} {\bibfnamefont {M.}~\bibnamefont
  {Hentschel}}, \bibinfo {author} {\bibfnamefont {R.}~\bibnamefont
  {Kienberger}}, \bibinfo {author} {\bibfnamefont {C.}~\bibnamefont
  {Spielmann}}, \bibinfo {author} {\bibfnamefont {G.~A.}\ \bibnamefont
  {Reider}}, \bibinfo {author} {\bibfnamefont {N.}~\bibnamefont {Milosevic}},
  \bibinfo {author} {\bibfnamefont {T.}~\bibnamefont {Brabec}}, \bibinfo
  {author} {\bibfnamefont {P.}~\bibnamefont {Corkum}}, \bibinfo {author}
  {\bibfnamefont {U.}~\bibnamefont {Heinzmann}}, \bibinfo {author}
  {\bibfnamefont {M.}~\bibnamefont {Drescher}}, \ and\ \bibinfo {author}
  {\bibfnamefont {F.}~\bibnamefont {Krausz}},\ }\href {\doibase
  10.1038/35107000} {\bibfield  {journal} {\bibinfo  {journal} {Nature}\
  }\textbf {\bibinfo {volume} {414}},\ \bibinfo {pages} {509} (\bibinfo {year}
  {2001})}\BibitemShut {NoStop}%
\bibitem [{\citenamefont {Drescher}\ \emph {et~al.}(2001)\citenamefont
  {Drescher}, \citenamefont {Hentschel}, \citenamefont {Kienberger},
  \citenamefont {Tempea}, \citenamefont {Spielmann}, \citenamefont {Reider},
  \citenamefont {Corkum},\ and\ \citenamefont {Krausz}}]{DreHenKie2001}%
  \BibitemOpen
  \bibfield  {author} {\bibinfo {author} {\bibfnamefont {M.}~\bibnamefont
  {Drescher}}, \bibinfo {author} {\bibfnamefont {M.}~\bibnamefont {Hentschel}},
  \bibinfo {author} {\bibfnamefont {R.}~\bibnamefont {Kienberger}}, \bibinfo
  {author} {\bibfnamefont {G.}~\bibnamefont {Tempea}}, \bibinfo {author}
  {\bibfnamefont {C.}~\bibnamefont {Spielmann}}, \bibinfo {author}
  {\bibfnamefont {G.~A.}\ \bibnamefont {Reider}}, \bibinfo {author}
  {\bibfnamefont {P.~B.}\ \bibnamefont {Corkum}}, \ and\ \bibinfo {author}
  {\bibfnamefont {F.}~\bibnamefont {Krausz}},\ }\href {\doibase
  10.1126/science.1058561} {\bibfield  {journal} {\bibinfo  {journal}
  {Science}\ }\textbf {\bibinfo {volume} {291}},\ \bibinfo {pages} {1923}
  (\bibinfo {year} {2001})}\BibitemShut {NoStop}%
\bibitem [{\citenamefont {Krausz}\ and\ \citenamefont
  {Ivanov}(2009)}]{KraIva2009}%
  \BibitemOpen
  \bibfield  {author} {\bibinfo {author} {\bibfnamefont {F.}~\bibnamefont
  {Krausz}}\ and\ \bibinfo {author} {\bibfnamefont {M.}~\bibnamefont
  {Ivanov}},\ }\href {\doibase 10.1103/RevModPhys.81.163} {\bibfield  {journal}
  {\bibinfo  {journal} {Rev. Mod. Phys}\ }\textbf {\bibinfo {volume} {81}},\
  \bibinfo {pages} {163} (\bibinfo {year} {2009})}\BibitemShut {NoStop}%
\bibitem [{Note1()}]{Note1}%
  \BibitemOpen
  \bibinfo {note} {The angular dependent differential cross section has been
  obtained from the sigma database at the \protect \href
  {http://www.nndc.bnl.gov/}{National Nuclear Database Center
  (NNDC)}.}\BibitemShut {Stop}%
\bibitem [{\citenamefont {Feist}\ \emph {et~al.}(2008)\citenamefont {Feist},
  \citenamefont {Nagele}, \citenamefont {Pazourek}, \citenamefont {Persson},
  \citenamefont {Schneider}, \citenamefont {Collins},\ and\ \citenamefont
  {Burgd\"{o}rfer}}]{FeiNagPaz2008}%
  \BibitemOpen
  \bibfield  {author} {\bibinfo {author} {\bibfnamefont {J.}~\bibnamefont
  {Feist}}, \bibinfo {author} {\bibfnamefont {S.}~\bibnamefont {Nagele}},
  \bibinfo {author} {\bibfnamefont {R.}~\bibnamefont {Pazourek}}, \bibinfo
  {author} {\bibfnamefont {E.}~\bibnamefont {Persson}}, \bibinfo {author}
  {\bibfnamefont {B.~I.}\ \bibnamefont {Schneider}}, \bibinfo {author}
  {\bibfnamefont {L.~A.}\ \bibnamefont {Collins}}, \ and\ \bibinfo {author}
  {\bibfnamefont {J.}~\bibnamefont {Burgd\"{o}rfer}},\ }\href {\doibase
  10.1103/PhysRevA.77.043420} {\bibfield  {journal} {\bibinfo  {journal} {Phys.
  Rev. A}\ }\textbf {\bibinfo {volume} {77}},\ \bibinfo {pages} {043420}
  (\bibinfo {year} {2008})}\BibitemShut {NoStop}%
\bibitem [{\citenamefont {Feist}\ \emph {et~al.}(2009)\citenamefont {Feist},
  \citenamefont {Nagele}, \citenamefont {Pazourek}, \citenamefont {Persson},
  \citenamefont {Schneider}, \citenamefont {Collins},\ and\ \citenamefont
  {Burgd\"{o}rfer}}]{FeiNagPaz2009}%
  \BibitemOpen
  \bibfield  {author} {\bibinfo {author} {\bibfnamefont {J.}~\bibnamefont
  {Feist}}, \bibinfo {author} {\bibfnamefont {S.}~\bibnamefont {Nagele}},
  \bibinfo {author} {\bibfnamefont {R.}~\bibnamefont {Pazourek}}, \bibinfo
  {author} {\bibfnamefont {E.}~\bibnamefont {Persson}}, \bibinfo {author}
  {\bibfnamefont {B.~I.}\ \bibnamefont {Schneider}}, \bibinfo {author}
  {\bibfnamefont {L.~A.}\ \bibnamefont {Collins}}, \ and\ \bibinfo {author}
  {\bibfnamefont {J.}~\bibnamefont {Burgd\"{o}rfer}},\ }\href {\doibase
  10.1103/PhysRevLett.103.063002} {\bibfield  {journal} {\bibinfo  {journal}
  {Phys. Rev. Lett.}\ }\textbf {\bibinfo {volume} {103}},\ \bibinfo {pages}
  {063002} (\bibinfo {year} {2009})}\BibitemShut {NoStop}%
\bibitem [{\citenamefont {Park}\ and\ \citenamefont
  {Light}(1986)}]{ParkLight86}%
  \BibitemOpen
  \bibfield  {author} {\bibinfo {author} {\bibfnamefont {T.~J.}\ \bibnamefont
  {Park}}\ and\ \bibinfo {author} {\bibfnamefont {J.~C.}\ \bibnamefont
  {Light}},\ }\href {\doibase 10.1063/1.451548} {\bibfield  {journal} {\bibinfo
   {journal} {J. Chem. Phys.}\ }\textbf {\bibinfo {volume} {85}},\ \bibinfo
  {pages} {5870} (\bibinfo {year} {1986})}\BibitemShut {NoStop}%
\bibitem [{\citenamefont {Palacios}\ \emph {et~al.}(2007)\citenamefont
  {Palacios}, \citenamefont {McCurdy},\ and\ \citenamefont
  {Rescigno}}]{PalMccRes2007}%
  \BibitemOpen
  \bibfield  {author} {\bibinfo {author} {\bibfnamefont {A.}~\bibnamefont
  {Palacios}}, \bibinfo {author} {\bibfnamefont {C.~W.}\ \bibnamefont
  {McCurdy}}, \ and\ \bibinfo {author} {\bibfnamefont {T.~N.}\ \bibnamefont
  {Rescigno}},\ }\href {\doibase 10.1103/PhysRevA.76.043420} {\bibfield
  {journal} {\bibinfo  {journal} {Phys. Rev. A}\ }\textbf {\bibinfo {volume}
  {76}},\ \bibinfo {pages} {043420} (\bibinfo {year} {2007})}\BibitemShut
  {NoStop}%
\bibitem [{\citenamefont {Burgd\"{o}rfer}\ \emph {et~al.}(1997)\citenamefont
  {Burgd\"{o}rfer}, \citenamefont {Qiu}, \citenamefont {Wang},\ and\
  \citenamefont {McGuire}}]{BurQiuWan1997}%
  \BibitemOpen
  \bibfield  {author} {\bibinfo {author} {\bibfnamefont {J.}~\bibnamefont
  {Burgd\"{o}rfer}}, \bibinfo {author} {\bibfnamefont {Y.}~\bibnamefont {Qiu}},
  \bibinfo {author} {\bibfnamefont {J.}~\bibnamefont {Wang}}, \ and\ \bibinfo
  {author} {\bibfnamefont {J.~H.}\ \bibnamefont {McGuire}},\ }\href {\doibase
  10.1063/1.52257} {\bibfield  {journal} {\bibinfo  {journal} {AIP Conference
  Proceedings}\ }\textbf {\bibinfo {volume} {389}},\ \bibinfo {pages} {475}
  (\bibinfo {year} {1997})}\BibitemShut {NoStop}%
\bibitem [{\citenamefont {McGuire}\ \emph {et~al.}(1995)\citenamefont
  {McGuire}, \citenamefont {Berrah}, \citenamefont {Bartlett}, \citenamefont
  {Samson}, \citenamefont {Tanis}, \citenamefont {Cocke},\ and\ \citenamefont
  {Schlachter}}]{McGBerBar1995}%
  \BibitemOpen
  \bibfield  {author} {\bibinfo {author} {\bibfnamefont {J.~H.}\ \bibnamefont
  {McGuire}}, \bibinfo {author} {\bibfnamefont {N.}~\bibnamefont {Berrah}},
  \bibinfo {author} {\bibfnamefont {R.~J.}\ \bibnamefont {Bartlett}}, \bibinfo
  {author} {\bibfnamefont {J.~A.~R.}\ \bibnamefont {Samson}}, \bibinfo {author}
  {\bibfnamefont {J.~A.}\ \bibnamefont {Tanis}}, \bibinfo {author}
  {\bibfnamefont {C.~L.}\ \bibnamefont {Cocke}}, \ and\ \bibinfo {author}
  {\bibfnamefont {A.~S.}\ \bibnamefont {Schlachter}},\ }\href {\doibase
  10.1088/0953-4075/28/6/009} {\bibfield  {journal} {\bibinfo  {journal} {J.
  Phys. B}\ }\textbf {\bibinfo {volume} {28}},\ \bibinfo {pages} {913}
  (\bibinfo {year} {1995})}\BibitemShut {NoStop}%
\bibitem [{\citenamefont {Byron}\ and\ \citenamefont
  {Joachain}(1967)}]{ByrJoa1967}%
  \BibitemOpen
  \bibfield  {author} {\bibinfo {author} {\bibfnamefont {F.~W.}\ \bibnamefont
  {Byron}}\ and\ \bibinfo {author} {\bibfnamefont {C.~J.}\ \bibnamefont
  {Joachain}},\ }\href {\doibase 10.1103/PhysRev.164.1} {\bibfield  {journal}
  {\bibinfo  {journal} {Phys. Rev.}\ }\textbf {\bibinfo {volume} {164}},\
  \bibinfo {pages} {1} (\bibinfo {year} {1967})}\BibitemShut {NoStop}%
\bibitem [{\citenamefont {\r{A}berg}(1970)}]{Abe1970}%
  \BibitemOpen
  \bibfield  {author} {\bibinfo {author} {\bibfnamefont {T.}~\bibnamefont
  {\r{A}berg}},\ }\href {\doibase 10.1103/PhysRevA.2.1726} {\bibfield
  {journal} {\bibinfo  {journal} {Phys. Rev. A}\ }\textbf {\bibinfo {volume}
  {2}},\ \bibinfo {pages} {1726} (\bibinfo {year} {1970})}\BibitemShut
  {NoStop}%
\bibitem [{\citenamefont {Dalgarno}\ and\ \citenamefont
  {Sadeghpour}(1992)}]{DalSad1992}%
  \BibitemOpen
  \bibfield  {author} {\bibinfo {author} {\bibfnamefont {A.}~\bibnamefont
  {Dalgarno}}\ and\ \bibinfo {author} {\bibfnamefont {H.~R.}\ \bibnamefont
  {Sadeghpour}},\ }\href {\doibase 10.1103/PhysRevA.46.R3591} {\bibfield
  {journal} {\bibinfo  {journal} {Phys. Rev. A}\ }\textbf {\bibinfo {volume}
  {46}},\ \bibinfo {pages} {R3591} (\bibinfo {year} {1992})}\BibitemShut
  {NoStop}%
\bibitem [{\citenamefont {Andersson}\ and\ \citenamefont
  {Burgd\"{o}rfer}(1993)}]{AndBur1993}%
  \BibitemOpen
  \bibfield  {author} {\bibinfo {author} {\bibfnamefont {L.~R.}\ \bibnamefont
  {Andersson}}\ and\ \bibinfo {author} {\bibfnamefont {J.}~\bibnamefont
  {Burgd\"{o}rfer}},\ }\href {\doibase 10.1103/PhysRevLett.71.50} {\bibfield
  {journal} {\bibinfo  {journal} {Phys. Rev. Lett.}\ }\textbf {\bibinfo
  {volume} {71}},\ \bibinfo {pages} {50} (\bibinfo {year} {1993})}\BibitemShut
  {NoStop}%
\bibitem [{\citenamefont {Andersson}\ and\ \citenamefont
  {Burgd\"{o}rfer}(1994)}]{AndBur1994}%
  \BibitemOpen
  \bibfield  {author} {\bibinfo {author} {\bibfnamefont {L.~R.}\ \bibnamefont
  {Andersson}}\ and\ \bibinfo {author} {\bibfnamefont {J.}~\bibnamefont
  {Burgd\"{o}rfer}},\ }\href {\doibase 10.1103/PhysRevA.50.R2810} {\bibfield
  {journal} {\bibinfo  {journal} {Phys. Rev. A}\ }\textbf {\bibinfo {volume}
  {50}},\ \bibinfo {pages} {R2810} (\bibinfo {year} {1994})}\BibitemShut
  {NoStop}%
\bibitem [{\citenamefont {Suri\'{c}}\ \emph {et~al.}(1994)\citenamefont
  {Suri\'{c}}, \citenamefont {Pisk}, \citenamefont {Logan},\ and\ \citenamefont
  {Pratt}}]{SurLogPra1994}%
  \BibitemOpen
  \bibfield  {author} {\bibinfo {author} {\bibfnamefont {T.}~\bibnamefont
  {Suri\'{c}}}, \bibinfo {author} {\bibfnamefont {K.}~\bibnamefont {Pisk}},
  \bibinfo {author} {\bibfnamefont {B.~A.}\ \bibnamefont {Logan}}, \ and\
  \bibinfo {author} {\bibfnamefont {R.~H.}\ \bibnamefont {Pratt}},\ }\href
  {\doibase 10.1103/PhysRevLett.73.790} {\bibfield  {journal} {\bibinfo
  {journal} {Phys. Rev. Lett.}\ }\textbf {\bibinfo {volume} {73}},\ \bibinfo
  {pages} {790} (\bibinfo {year} {1994})}\BibitemShut {NoStop}%
\bibitem [{\citenamefont {Andersen}\ \emph {et~al.}(1987)\citenamefont
  {Andersen}, \citenamefont {Hvelplund}, \citenamefont {Knudsen}, \citenamefont
  {Moller}, \citenamefont {Sorensen}, \citenamefont {Elsener}, \citenamefont
  {Rensfelt},\ and\ \citenamefont {Uggerhoj}}]{AndHveKnu1987}%
  \BibitemOpen
  \bibfield  {author} {\bibinfo {author} {\bibfnamefont {L.~H.}\ \bibnamefont
  {Andersen}}, \bibinfo {author} {\bibfnamefont {P.}~\bibnamefont {Hvelplund}},
  \bibinfo {author} {\bibfnamefont {H.}~\bibnamefont {Knudsen}}, \bibinfo
  {author} {\bibfnamefont {S.~P.}\ \bibnamefont {Moller}}, \bibinfo {author}
  {\bibfnamefont {A.~H.}\ \bibnamefont {Sorensen}}, \bibinfo {author}
  {\bibfnamefont {K.}~\bibnamefont {Elsener}}, \bibinfo {author} {\bibfnamefont
  {K.~G.}\ \bibnamefont {Rensfelt}}, \ and\ \bibinfo {author} {\bibfnamefont
  {E.}~\bibnamefont {Uggerhoj}},\ }\href {\doibase 10.1103/PhysRevA.36.3612}
  {\bibfield  {journal} {\bibinfo  {journal} {Phys. Rev. A}\ }\textbf {\bibinfo
  {volume} {36}},\ \bibinfo {pages} {3612} (\bibinfo {year}
  {1987})}\BibitemShut {NoStop}%
\bibitem [{\citenamefont {deHarak}\ \emph {et~al.}(2006)\citenamefont
  {deHarak}, \citenamefont {Childers},\ and\ \citenamefont
  {Martin}}]{HarChiMar2006}%
  \BibitemOpen
  \bibfield  {author} {\bibinfo {author} {\bibfnamefont {B.~A.}\ \bibnamefont
  {deHarak}}, \bibinfo {author} {\bibfnamefont {J.~G.}\ \bibnamefont
  {Childers}}, \ and\ \bibinfo {author} {\bibfnamefont {N.~L.~S.}\ \bibnamefont
  {Martin}},\ }\href {\doibase 10.1103/PhysRevA.74.032714} {\bibfield
  {journal} {\bibinfo  {journal} {Phys. Rev. A}\ }\textbf {\bibinfo {volume}
  {74}},\ \bibinfo {pages} {032714} (\bibinfo {year} {2006})}\BibitemShut
  {NoStop}%
\bibitem [{\citenamefont {Moshammer}\ \emph {et~al.}(2007)\citenamefont
  {Moshammer}, \citenamefont {Jiang}, \citenamefont {Foucar}, \citenamefont
  {Rudenko}, \citenamefont {Ergler}, \citenamefont {Schr\"{o}ter},
  \citenamefont {L\"{u}demann}, \citenamefont {Zrost}, \citenamefont {Fischer},
  \citenamefont {Titze}, \citenamefont {Jahnke}, \citenamefont {Sch\"{o}ffler},
  \citenamefont {Weber}, \citenamefont {D\"{o}rner}, \citenamefont {Zouros},
  \citenamefont {Dorn}, \citenamefont {Ferger}, \citenamefont {K\"{u}hnel},
  \citenamefont {D\"{u}sterer}, \citenamefont {Treusch}, \citenamefont
  {Radcliffe}, \citenamefont {Pl\"{o}njes},\ and\ \citenamefont
  {Ullrich}}]{MosJiaFou2007}%
  \BibitemOpen
  \bibfield  {author} {\bibinfo {author} {\bibfnamefont {R.}~\bibnamefont
  {Moshammer}}, \bibinfo {author} {\bibfnamefont {Y.~H.}\ \bibnamefont
  {Jiang}}, \bibinfo {author} {\bibfnamefont {L.}~\bibnamefont {Foucar}},
  \bibinfo {author} {\bibfnamefont {A.}~\bibnamefont {Rudenko}}, \bibinfo
  {author} {\bibfnamefont {T.}~\bibnamefont {Ergler}}, \bibinfo {author}
  {\bibfnamefont {C.~D.}\ \bibnamefont {Schr\"{o}ter}}, \bibinfo {author}
  {\bibfnamefont {S.}~\bibnamefont {L\"{u}demann}}, \bibinfo {author}
  {\bibfnamefont {K.}~\bibnamefont {Zrost}}, \bibinfo {author} {\bibfnamefont
  {D.}~\bibnamefont {Fischer}}, \bibinfo {author} {\bibfnamefont
  {J.}~\bibnamefont {Titze}}, \bibinfo {author} {\bibfnamefont
  {T.}~\bibnamefont {Jahnke}}, \bibinfo {author} {\bibfnamefont
  {M.}~\bibnamefont {Sch\"{o}ffler}}, \bibinfo {author} {\bibfnamefont
  {T.}~\bibnamefont {Weber}}, \bibinfo {author} {\bibfnamefont
  {R.}~\bibnamefont {D\"{o}rner}}, \bibinfo {author} {\bibfnamefont {T.~J.~M.}\
  \bibnamefont {Zouros}}, \bibinfo {author} {\bibfnamefont {A.}~\bibnamefont
  {Dorn}}, \bibinfo {author} {\bibfnamefont {T.}~\bibnamefont {Ferger}},
  \bibinfo {author} {\bibfnamefont {K.~U.}\ \bibnamefont {K\"{u}hnel}},
  \bibinfo {author} {\bibfnamefont {S.}~\bibnamefont {D\"{u}sterer}}, \bibinfo
  {author} {\bibfnamefont {R.}~\bibnamefont {Treusch}}, \bibinfo {author}
  {\bibfnamefont {P.}~\bibnamefont {Radcliffe}}, \bibinfo {author}
  {\bibfnamefont {E.}~\bibnamefont {Pl\"{o}njes}}, \ and\ \bibinfo {author}
  {\bibfnamefont {J.}~\bibnamefont {Ullrich}},\ }\href {\doibase
  10.1103/PhysRevLett.98.203001} {\bibfield  {journal} {\bibinfo  {journal}
  {Phys. Rev. Lett.}\ }\textbf {\bibinfo {volume} {98}},\ \bibinfo {pages}
  {203001} (\bibinfo {year} {2007})}\BibitemShut {NoStop}%
\bibitem [{\citenamefont {Sorokin}\ \emph {et~al.}(2007)\citenamefont
  {Sorokin}, \citenamefont {Wellhofer}, \citenamefont {Bobashev}, \citenamefont
  {Tiedtke},\ and\ \citenamefont {Richter}}]{SorWelBob2007}%
  \BibitemOpen
  \bibfield  {author} {\bibinfo {author} {\bibfnamefont {A.~A.}\ \bibnamefont
  {Sorokin}}, \bibinfo {author} {\bibfnamefont {M.}~\bibnamefont {Wellhofer}},
  \bibinfo {author} {\bibfnamefont {S.~V.}\ \bibnamefont {Bobashev}}, \bibinfo
  {author} {\bibfnamefont {K.}~\bibnamefont {Tiedtke}}, \ and\ \bibinfo
  {author} {\bibfnamefont {M.}~\bibnamefont {Richter}},\ }\href {\doibase
  10.1103/PhysRevA.75.051402} {\bibfield  {journal} {\bibinfo  {journal} {Phys.
  Rev. A}\ }\textbf {\bibinfo {volume} {75}},\ \bibinfo {pages} {051402(R)}
  (\bibinfo {year} {2007})}\BibitemShut {NoStop}%
\bibitem [{\citenamefont {Beutler}(1935)}]{Beutler35}%
  \BibitemOpen
  \bibfield  {author} {\bibinfo {author} {\bibfnamefont {H.}~\bibnamefont
  {Beutler}},\ }\href {\doibase 10.1007/BF01365116} {\bibfield  {journal}
  {\bibinfo  {journal} {Z. Phys. A}\ }\textbf {\bibinfo {volume} {93}},\
  \bibinfo {pages} {177} (\bibinfo {year} {1935})}\BibitemShut {NoStop}%
\bibitem [{\citenamefont {Fano}(1935)}]{Fano35}%
  \BibitemOpen
  \bibfield  {author} {\bibinfo {author} {\bibfnamefont {U.}~\bibnamefont
  {Fano}},\ }\href {\doibase 10.1007/BF02958288} {\bibfield  {journal}
  {\bibinfo  {journal} {Nuovo Cimento (1924-1942)}\ }\textbf {\bibinfo {volume}
  {12}},\ \bibinfo {pages} {154} (\bibinfo {year} {1935})}\BibitemShut
  {NoStop}%
\bibitem [{\citenamefont {Fano}(1961)}]{Fano61}%
  \BibitemOpen
  \bibfield  {author} {\bibinfo {author} {\bibfnamefont {U.}~\bibnamefont
  {Fano}},\ }\href {\doibase 10.1103/PhysRev.124.1866} {\bibfield  {journal}
  {\bibinfo  {journal} {Phys. Rev.}\ }\textbf {\bibinfo {volume} {124}},\
  \bibinfo {pages} {1866} (\bibinfo {year} {1961})}\BibitemShut {NoStop}%
\bibitem [{\citenamefont {Sinano\u{g}lu}\ and\ \citenamefont
  {Herrick}(1975)}]{SinHer1973}%
  \BibitemOpen
  \bibfield  {author} {\bibinfo {author} {\bibfnamefont {O.}~\bibnamefont
  {Sinano\u{g}lu}}\ and\ \bibinfo {author} {\bibfnamefont {D.~R.}\ \bibnamefont
  {Herrick}},\ }\href {\doibase 10.1063/1.430540} {\bibfield  {journal}
  {\bibinfo  {journal} {J. Chem. Phys.}\ }\textbf {\bibinfo {volume} {62}},\
  \bibinfo {pages} {886} (\bibinfo {year} {1975})}\BibitemShut {NoStop}%
\end{thebibliography}
\end{document}